\documentclass[%
 reprint,
 amsmath,amssymb,
 aps,
]{revtex4-2}

\usepackage{graphicx}
\usepackage{dcolumn}
\usepackage{bm}
\usepackage{longtable}

\begin{document}

\preprint{APS/123-QED}

\title{Rigid triaxiality has the SU(3) symmetry: $^{166}$Er as an example}

\author{Chunxiao Zhou}
\email{zhouchunxiao567@163.com}
\author{Xue Shang}
\affiliation{College of Mathematics and Physics Science, Hunan University of Arts and Science, Changde 415000, People's Republic of China}

\author{Tao Wang}
\email{suiyueqiaoqiao@163.com}
\affiliation{College of Physics, Tonghua Normal University, Tonghua 134000, People's Republic of China}

\date{\today}

\begin{abstract}
The emergence of triaxiality in the low-lying collective bands of $^{166}$Er is systematically explored within the SU3-IBM. In this framework, SU(3) higher-order interactions are included, which enable the descriptions of various quadrupole deformations. The triaxiality of $^{166}$Er is described with a triaxiality angle $\gamma=9.7^{\circ}$. In addition, the calculated energy spectra, $B(E2)$ transition strengths, and quadrupole moments show excellent agreement with experimental data. These results provide further evidence supporting the SU(3) triaxial interpretation of $^{166}$Er and confirm its triaxial deformation rather than the prolate shape.

\end{abstract}

\maketitle


\section{INTRODUCTION}

The study of rotational dynamics has played a special role in the exploration of nuclear structure. The rotational bands emerge only in deformed nuclei, because rotation about a symmetry axis leaves the quantum state unchanged. In the traditional view, as emphasized by Aage Bohr in his Nobel lecture using $^{166}$Er as an example, a large fraction of heavy nuclei adopt axially symmetric prolate ellipsoids and rotate about one of the short axes \cite{Bohr}. However, recent experimental findings present alternative perspectives. Nuclei such as $^{238}$U \cite{Abdulhamid24} and $^{154}$Sm \cite{Kleeman25}, which were previously acknowledged for their axially symmetric shapes, have been observed to exhibit triaxial deformation. Meanwhile, Otsuka and collaborators demonstrated that the low-lying $\gamma$ bands in heavy strongly deformed nuclei, such as $^{166}$Er and $^{162}$Dy, are rotational excitations of triaxially deformed states rather than vibrational excitations as traditionally interpreted \cite{otsuka2019, tsunoda2023, otsuka2025, tsunodaa2025}. These experimental and theoretical findings suggest that triaxial shapes may occur in a large number of heavy deformed nuclei, and triaxiality plays a much more prominent and critical role than previously recognized.  

Over the years, numerous theoretical approaches have been employed to describe the nuclear shapes. In the geometric framework of Bohr-Mottelson model \cite{Bohr1975}, nuclear shape deformations are characterized by the quadrupole deformation parameters $\beta$  and $\gamma$. Typically, $\gamma = 0^{\circ}$ and $\gamma = 60^{\circ}$ correspond to the prolate and oblate axially symmetric shapes, respectively, while $0<\gamma<60^{\circ}$ label a triaxial rotor. The quadrupole deformation parameters $\beta$ and $\gamma$ can be recovered in the algebraic interacting boson model (IBM) \cite{Iachello75, iachello1987} by mapping its Hamiltonian onto a potential energy surface $E(\beta,\gamma)$ using the coherent-state formalism. In its simplest version, known as IBM-1, no distinction is made between bosons coming from proton pairs or neutron pairs, allowing a range of nuclear shapes to be effectively described. By now, it is well established that a two- and three-body IBM-1 Hamiltonian can give rise to axially symmetric shapes. The descriptions of triaxial shapes require higher-order interactions. Specifically, a stable triaxial minimum at $\gamma = 30^{\circ}$ on the potential energy surface can be induced by the cubic term $[d^{\dag} d^{\dag} d^{\dag}]^{(L)}\cdot[\tilde{d}\tilde{d}\tilde{d}]^{(L)}$ \cite{Isacker81,Isacker84}. Rigid triaxial rotation can be generated
by including higher-order interactions (up to four-body terms) in the SU(3) symmetry limit \cite{smirnov2000}. Alternatively, triaxiality can also be investigated using a mean field-derived IBM-1 Hamiltonian with an intrinsic triaxial deformation derived from fermionic
proxy-SU(3) irreducible representations \cite{vasileiou2025}.

The IBM-1 provides an algebraic language for describing quadrupole collectivity, but its quantitative description is inadequate for certain phenomena.
For example, a cluster of neutron-deficient nuclei, such as $^{168,170}$Os, $^{166}$W, $^{172}$Pt \cite{grahn2016, sayg2017, Cederwall18, goasduff2019}, has been experimentally found to exhibit a small B(E2) ratio  $B_{4/2} = B(E2; 4_{1}^{+}\rightarrow2_{1}^{+})/B(E2; 2_{1}^{+}\rightarrow0_{1}^{+})<1.0$. In the standard collective model, however, the B(E2) values increase with spin along the yrast band and thus the ratio $B_{4/2}$ is strictly greater than unity. Notably, this anomalous phenomenon cannot be reproduced in a very 
convincing way with large-scale shell model approaches \cite{Cederwall18} or self-consistent mean-field calculations \cite{ goasduff2019}. On the other hand, experimental data on Cd nuclei have called into question the phonon mode as a major paradigm for describing nuclei near closed shells (Cd puzzle) \cite{Garrett08, garrett2012, Heyde16, Garrett18}. It has been found that there is no $0_{3}^{+}$ state in the three-phonon levels of $^{120}$Cd \cite{batchelder2012}, and the observed $0_{3}^{+}$ state  in $^{112}$Cd has been interpreted as intruder excitation for a rotational band \cite{garrett2019}. These unusual experimental phenomena prompt us to revisit and enrich our understanding of the existing nuclear theories. 

In response to these challenges, an extension of the interacting boson model incorporating SU(3) higher-order interactions (hereafter denoted SU3-IBM) was recently proposed to address the two anomalous experimental phenomena \cite{Wang20, Wang22}. Distinct from the previous IBM-1, in this framework all deformations (prolate, oblate and triaxial) are governed by the SU(3) symmetry. Encouragingly, the SU3-IBM provides a consistent and compelling explanation for the aforementioned experimental anomalies \cite{Zhang22, Wangtao, Zhang24, Pan24, Zhang25, Zhang252, Teng25, Cheng25, Zhang253, Zhang254, Jin25, Li25, Wang25, WangPd106, Zhao25}. 
Beyond that, this model has also provided insightful interpretations of several other notable phenomena, such as the $\gamma$-soft character of $^{196}$Pt \cite{wang2024}, the prolate-oblate shape asymmetric transitions in the Hf-Hg region \cite{wang2023, Teng251}, E(5)-like spectra in $^{82}$Kr \cite{zhou2023}, and boson number odd-even effect in $^{196-204}$Hg \cite{WangHg}. These successes demonstrate the model's effectiveness and broad applicability. 

The ability of the SU3-IBM to describe triaxial nuclear shapes is particularly relevant here. Therefore, in this paper, the triaxiality of $^{166}$Er is investigated within the SU3-IBM framework. Based on the well-established correspondence between the SU(3) irreducible representations (irreps) and collective deformation parameters $(\beta, \gamma)$, the degree of triaxiality is systematically evaluated. In addition, our calculations also demonstrate that the energy spectra, B(E2) transition strengths and quadrupole moments fit well with the experimental results. These results provide further evidence reinforcing the triaxial interpretation of $^{166}$Er and confirm its SU(3) triaxial character over purely prolate configurations. A parallel analysis for $^{154}$Sm leads to similar conclusions \cite{zhou2026}.

\section{THE HAMILTONIAN OF the SU3-IBM}



The Hamiltonian used in this paper can be expressed as
\begin{equation}
\hat{H}=\alpha\hat{n}_{d}+\hat{H}_{Tri},
\label{eq1}
\end{equation}
where $\alpha$ is the fitting parameter, $\hat{n}_{d}=d^{\dag}\cdot\tilde{d}$ is the $d$-boson number operator in the U(5) symmetry limit. $\hat{H}_{
Tri}$ is rigid rotor Hamiltonian which can be divided into static and dynamic parts
\begin{equation}
\hat{H}_{Tri}=\hat{H}_{S}+\hat{H}_{D},
\label{eq2}
\end{equation}
where 
\begin{equation}
\hat{H}_{S}=-\frac{a_{1}}{2N}\hat{C}_{2}[\mathrm{SU(3)}]+\frac{a_{2}}{2N^{2}}\hat{C}_{3}[\mathrm{SU(3)}]+\frac{a_{3}}{2N^{3}}\hat{C}_{2}^{2}[\mathrm{SU(3)}], 
\label{eq3}
\end{equation}
\begin{equation}
\hat{H}_{D}=t_{1}[\hat{L}\times \hat{Q} \times \hat{L}]^{(0)}+t_{2}[(\hat{L}\times \hat{Q})^{(1)} \times (\hat{L} \times \hat{Q})^{(1)}]^{(0)}+t_{3}\hat{L}^{2},
\label{eq4}
\end{equation}
where $a_{1}$, $a_{2}$, $a_{3}$, $t_{1}$, $t_{2}$ and $t_{3}$ are the six fitting parameters.  $\hat{L}=\sqrt{10}[d^{\dag}\times\tilde{d}]^{1}$, $\hat{Q}=[d^{\dag}\times\tilde{s}+s^{\dag}\times \tilde{d}]^{(2)}-\frac{\sqrt{7}}{2}[d^{\dag}\times \tilde{d}]^{(2)} $ are the SU(3) generators. To realize a rigid triaxial rotor within this framework, the SU(3) limit of the Hamiltonian (\ref{eq2}) must contain higher-order terms \cite{smirnov2000}. Thus, Eq. (\ref{eq3}) includes the second-, third-, and fourth-order invariant operators of the $SU(3) \supset SO(3)$ integrity basis.
The SU(3) Casimir operators are defined as
\begin{equation}
\hat{C}_{2}[\mathrm{SU(3)}]=2\hat{Q}\cdot\hat{Q}+\frac{3}{4}\hat{L}^{2},
\label{eq5}
\end{equation}
\begin{equation}
\hat{C}_{3}[\mathrm{SU(3)}]=-\frac{4\sqrt{35}}{9}[\hat{Q}\times\hat{Q}\times\hat{Q}]^{(0)}-\frac{\sqrt{15}}{2}[\hat{L}\times\hat{Q}\times\hat{L}]^{(0)}.
\label{eq6}
\end{equation}
And the eigenvalues of the two SU(3) Casimir operators can be expressed in terms of the SU(3) irrep ($\lambda, \mu$)
\begin{equation}
\langle \hat{C}_{2}[\mathrm{SU(3)}] \rangle=\lambda^{2}+\mu^{2}+3\lambda+3\mu+\lambda\mu,
\label{eq7}
\end{equation}
\begin{equation}
\langle \hat{C}_{3}[\mathrm{SU(3)}] \rangle=\frac{1}{9}(\lambda-\mu)(2\lambda+\mu+3)(\lambda+2\mu+3).
\label{eq8}
\end{equation}

The ground-state energy of the rigid triaxial Hamiltonian (\ref{eq2}) is given by $E_{g} = \langle \hat{H}_{S}\rangle = f(\lambda,\mu)$ evaluated at the optimal values $(\lambda,\mu)$, which are themselves determined by the parameters $a_{1}$, $a_{2}$, $a_{3}$. The relationship between the parameters $a_{1}$, $a_{2}$ and $a_{3}$ and SU(3) irrep ($\lambda,\mu$) can be derived as \cite{zhou2023}
\begin{equation}
a_{3}=\frac{a_{1} N^{2}}{2g(\lambda,\mu)}-\frac{a_{2} N}{6g(\lambda,\mu)}(3+\lambda+2\mu ),
\label{eq10}
\end{equation}
where $g(\lambda,\mu)=\lambda^{2}+\mu^{2}+3\lambda+3\mu+\lambda\mu$. For any arbitrary SU(3) irrep $(\lambda, \mu)$, it can serve as the ground state irrep when the parameters in $\hat{H}_{S}$ satisfy the condition of Eq. (\ref{eq10}). The dynamical interaction $\hat{H}_{D}$ is the SU(3) image of a rigid triaxial rotor \cite{smirnov2000}. The terms of $\hat{H}_{D}$ do not contribute to the ground state energy or other $0^{+}$ states, but they can modify the properties of non-$0^{+}$ states.

In this Hamiltonian, the SU(3) symmetry governs all quadrupole deformations: the second-order Casimir operator $-\hat{C}_{2}$[SU(3)] generates the prolate shape, whereas the third-order Casimir operator $\hat{C}_{3}$[SU(3)] yields the oblate shape. The combinations of the square of the second-order Casimir operator $-\hat{C}^{2}_{2}$[SU(3)] with $-\hat{C}_{2}$[SU(3)] and $\hat{C}_{3}$[SU(3)] can lead to the emergence of triaxial shapes \cite{smirnov2000}. Furthermore, by combining the $d$ boson number operator with the rigid triaxial shapes (the coupling of $\hat{n}_{d}$ with $\hat{H}_{S}$), various $\gamma$-soft modes different from the traditional O(6) $\gamma$-soft description can emerge \cite{wang2024}. 

In this paper, we focus on the  triaxial deformation of $^{166}$Er, the value of $\gamma$ is one of the key quantities of interest. This value has been calculated by Otsuka \textit{et al.} using the Monte Carlo Shell Model (MCSM) and can alternatively be estimated within the framework of the Davydov-Filippov model. The latter model has recently been highlighted by Otsuka \textit{et al.}, who stated: "The concept of triaxial shapes, one of the two main messages by Davydov and his collaborators, is definitely appropriate from the present view, and deserves high appreciation". Within this model, $\gamma$ can be extracted from the ratio of $B(E2)$ transition probabilities:

\begin{equation}
    \frac{B(E2;2_{2}^{+}\rightarrow 0_{1}^{+})}{B(E2;2_{1}^{+}\rightarrow 0_{1}^{+})}=\frac{\sqrt{9-8\sin^{2}3\gamma}-3+2\sin^{2}3\gamma}{\sqrt{9-8\sin^{2}3\gamma}+3-2\sin^{2}3\gamma}.
    \label{eq12}
\end{equation}

Based on the relationship between the SU(3) irrep and the collective rotor parameters $(\beta, \gamma)$, $\gamma$ is connected with the SU(3) irrep $(\lambda, \mu)$ via the relation \cite{castanos1988}

\begin{equation}
    \gamma=\tan^{-1}\left[\frac{\sqrt{3}(\mu+1)}{2\lambda+\mu+3}\right].
    \label{eq9}
\end{equation}
Consequently, Eq. (\ref{eq9}) enables the inverse determination of the irrep $(\lambda, \mu)$ form a  determined $\gamma$ value. However, it should be noted that in our model, $\gamma$ value is not solely determined by a single SU(3) irrep $(\lambda, \mu)$, because the terms $\hat{n}_{d}$ and $\hat{H}_{D}$ in Hamiltonian (\ref{eq1}) can break the SU(3) symmetry, resulting in mixing among different SU(3) irreps. Therefore, in this paper, the $\gamma$ represents an effective mean value that can be calculated according to the relation \cite{pan2003, Wei2025}
\begin{equation}
\bar{\gamma}=\sum d_{(\lambda, \mu)}\gamma_{(\lambda, \mu)}, 
\label{eq11}
\end{equation}
where $d_{(\lambda,\mu)}$ denotes the decomposition probability of the SU(3) irrep $(\lambda,\mu)$ in the given state, and $\gamma_{(\lambda, \mu)}$ denotes the triaxial deformation parameter associated with the irrep $(\lambda,\mu)$ via Eq. (\ref{eq9}).

\section{The Calculation of $^{166}$$\mathrm{Er}$}

In Refs. \cite{otsuka2019, tsunoda2023, otsuka2025}, through quadrupole matrix element calculations, Otsuka \textit{et al.} demonstrated that $^{166}$Er exhibits a triaxial deformation with $\gamma=8.2^{\circ}$.  Analysis of experimental $B(E2)$ values within the Davydov-Filippov model using Eq. (\ref{eq12}) yield a comparable triaxiality of $\gamma = 9.1^{\circ}$. These consistent findings motivate a reinterpretation of the nuclear shape of $^{166}$Er - not as a simple prolate ellipsoid, but rather as a more intricately rotating triaxial system. For the present study of $^{166}$Er, with boson number $N = 15$ and $\gamma = 8.2^{\circ}$, the corresponding SU(3) irrep can be approximately obtained as (22,4) using Eq. (\ref{eq9}). The Hamiltonian parameters are subsequently fixed through the following procedure. First, based on the experimental energies of the ground states ($0_{2}^{+}$ and $0_{3}^{+}$) and Eq. (\ref{eq10}), the parameters of Hamiltonian $\hat{H}_{S}$ are determined as $a_{1}=-4016.6$ KeV, $a_{2}=1205.0$ KeV, $a_{3}=522.2$ KeV. Subsequently, $\alpha=167.4$ KeV is chosen to optimize the agreement between calculated and experimental ground-state energies. Next, using the experimental
values of $B(E2;2_{1}^{+}\rightarrow 0_{1}^{+})$ and $B(E2;2_{2}^{+}\rightarrow 0_{1}^{+})$, the parameters $t_{1} = 10.5$ KeV and $t_{2} = - 0.99$ KeV are determined. Finally, $t_{3} = 25.9$ KeV is chosen to improve the agreement between the calculated higher energy levels and the experimental results.

\begin{figure}[b]
\includegraphics[width=1.0\columnwidth]{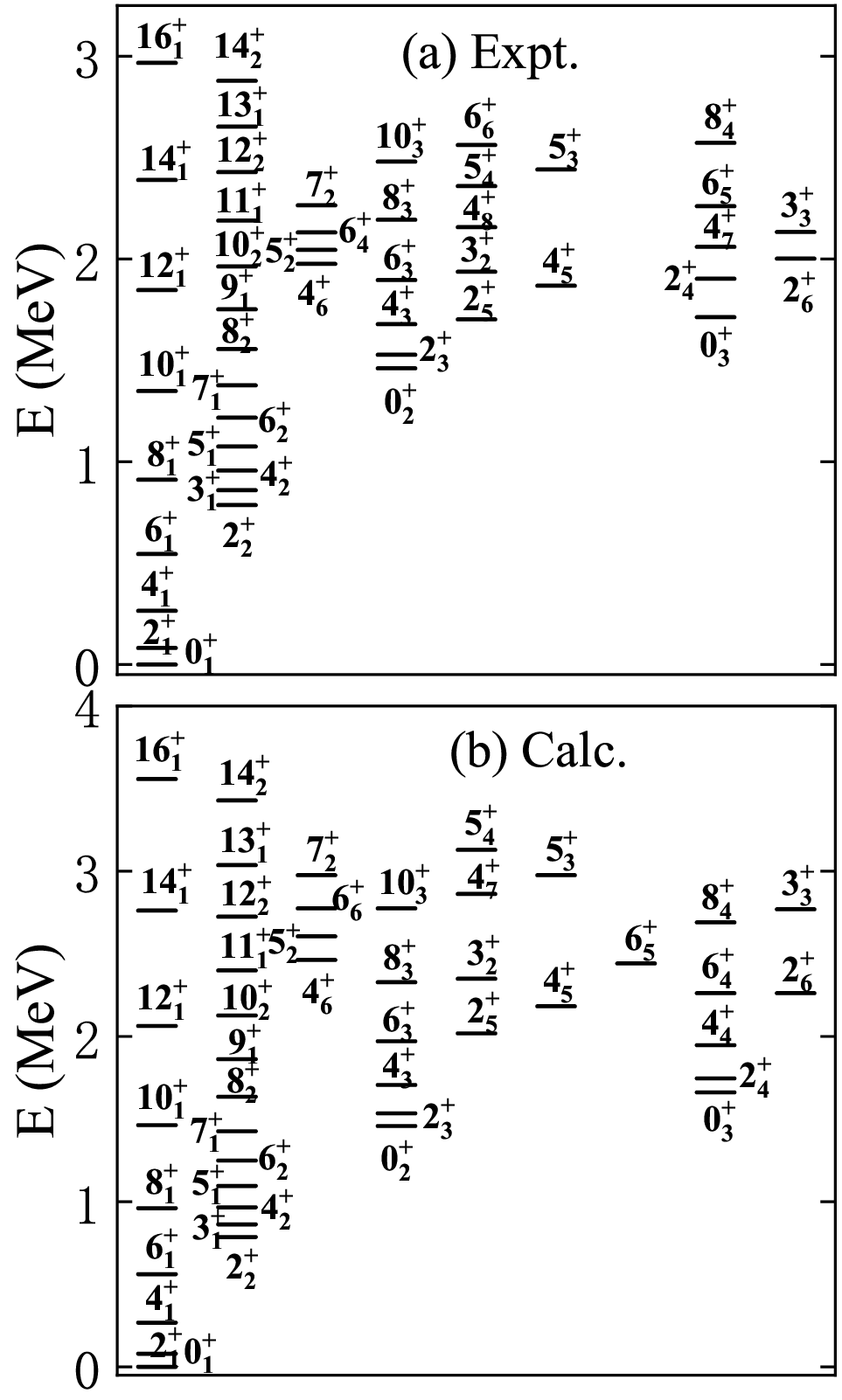}
\caption{\label{fig:epsart} Levels of $^{166}$Er. (a) experimental levels, (b) calculated levels based on the SU3-IBM Hamiltonian. The parameters used here are  $\alpha = 167.4$ KeV, $a_{1} = -4016.6$ KeV, $a_{2}=1205.0$ KeV, $a_{3}=522.2$ KeV, $t_{1}=10.5$ KeV, $t_{2}=-0.99$ KeV, $t_{3}=25.9$ KeV.}\label{fig1}
\end{figure}

 A comparison between the experimental (a) and calculated (b) energy level structures of $^{166}$Er is presented in Fig. \ref{fig1}. It can be seen that the ground-state band exhibits near-perfect agreement with the experimental data. The calculated $\gamma$-band also reproduces the experimental results well. In addition, we calculate the bands headed by the $0_{2}^{+}$ and $0_{3}^{+}$ states. The band headed by the $0_{2}^{+}$ state can well reproduce the experimental results, for example, the close proximity of the $0_{2}^{+}$ and $2_{3}^{+}$ states observed experimentally is perfectly reproduced by the calculation. For the band headed by the $0_{3}^{+}$ state, the calculated energy level $2_{4}^{+}$ is slightly lower than the experimental result, while other energy levels within this band show good consistency with measurements. Additionally, higher-lying bands built on the $2_{5}^{+}$, $4_{5}^{+}$ and $2_{6}^{+}$ states were also investigated theoretically, and the calculated level energies are similar to the experimental results. One notable difference between the experiment and our calculation is the absence of a well-defined $6^{+}$ state built on the $5_{4}^{+}$ state, whereas the calculated predicts a $6_{5}^{+}$ that is assigned to a distinct band.   
 Overall, apart from a minor overestimation of the higher-lying states, the theoretical results reproduce the experimental data satisfactorily. 

The existence of $\hat{n}_{d}$ and $\hat{H}_{D}$ breaks the SU(3) symmetry. Table \ref{Table1} presents the SU(3) decomposition of several low-lying states, each containing different irrep components. However, the dominant SU(3) irreps of the three lowest $0^{+}$ states remain the same as those obtained in the SU(3) limit. According to Eqs. (\ref{eq9}) and (\ref{eq11}), the $\gamma$ values of some low-lying states are listed in Table \ref{Table2}. It can be seen that our calculations are slightly larger than the results given in Ref. \cite{otsuka2025}. For states $0_{1}^{+}$, $2_{1}^{+}$, $2_{2}^{+}$, $3_{1}^{+}$, $4_{1}^{+}$, and $4_{6}^{+}$, the dominant SU(3) irrep is (22,4). These states can be meaningfully characterized by Elliott's quantum number $K$. As shown in Fig. (\ref{fig1}), the ground band with $K=0$ contains the $0_{1}^{+}$, $2_{1}^{+}$, $4_{1}^{+}$ states, all of which correspond to nearly the same $\gamma$ value. The $\gamma$ band with $K=2$ contains $2_{2}^{+}$ and $3_{1}^{+}$ sates, corresponding to $\gamma=9.9^{\circ}$. The band with $K=4$ is build on the $4_{6}^{+}$ state, corresponding to $\gamma=10^{\circ}$. A similar correspondence can be found in Ref \cite{otsuka2025}. 
Our theory also predicts the $\gamma$ values of $0_{2}^{+}$ and $0_{3}^{+}$ states, and experimental verification is anticipated. 

\begin{table}
\centering 
\setlength{\tabcolsep}{0.9mm}
\caption{SU(3) decomposition of some low-lying states.}
\begin{tabular}{l c c c c c c c c c}
 \hline
 \hline
 \quad  & $0_{1}^{+}$ & $0_{2}^{+}$ & $0_{3}^{+}$& $2_{1}^{+}$& $2_{2}^{+}$ & $3_{1}^{+}$ & $4_{1}^{+}$ & $4_{6}^{+}$  \\ 
 \hline
 (26,2) &0.044 & 0.322& 0.001& 0.044 & 0.025 & 0.025 & 0.044 & \quad\\ 
 (22,4) & \textbf{0.882} & 0.011& 0.080 & \textbf{0.883} & \textbf{0.907} & \textbf{0.909} & \textbf{0.887} & \textbf{0.936} \\ 
 (18,6) & 0.059 & 0.155 & \textbf{0.73} & 0.057 & 0.065 & 0.063 & 0.053 & 0.061 \\ 
 (14,8) & \quad & 0.003& 0.017 & \quad & \quad & \quad & \quad & \quad \\ 
 (24,0) & 0.011 & \textbf{0.503}& 0.165 & 0.011 & \quad & \quad & 0.011 & \quad \\ 
 (20,2) & 0.004 & 0.005& 0.006 & 0.004 & 0.002 & 0.002 & 0.004 & 0.002 \\ 
 \hline
 \hline
\end{tabular}
\label{Table1}
\end{table}

\begin{table}
\centering 
\setlength{\tabcolsep}{5.0mm}
\caption{The deformation parameters $\gamma$ of some low-lying states in $^{166}$Er obtained from different approaches. The MSCM data are taken from Ref. \cite{otsuka2025}. The SU3-IBM results are obtained according to Eq. (\ref{eq11}).}
\begin{tabular}{l c c c} 
 \hline
 \hline
 \quad &  MCSM & SU3-IBM \\ 
 \hline
 $0_{1}^{+}$  & 8.2$^{\circ}$  & 9.7$^{\circ}$  \\
 $0_{2}^{+}$  & \quad & 5.2$^{\circ}$  \\

 $0_{3}^{+}$  & \quad & 12.5$^{\circ}$  \\ 
 $2_{1}^{+}$  & 8.2$^{\circ}$  & 9.6$^{\circ}$  \\
 $2_{2}^{+}$  & 9.1$^{\circ}$  & 9.9$^{\circ}$  \\
 
 $3_{1}^{+}$  & 9.1$^{\circ}$  & 9.9$^{\circ}$  \\
 
 $4_{1}^{+}$  & 8.2$^{\circ}$  & 9.6$^{\circ}$  \\ 
 $4_{6}^{+}$  & 9.5$^{\circ}$  & 10$^{\circ}$  \\ 
 \hline
 \hline
\end{tabular}
\label{Table2}
\end{table}

\begin{table}
 \renewcommand\arraystretch{1.15}
\caption{\label{tab:table3}%
Absolute $B(E2)$ values in W.u. for $E2$ transitions between the low-lying normal states in $^{166}$Er. Experimental values are taken from Ref. \cite{datafile, Bucurescu12019}. Theoretical calculations are carried out using the same parameters as these used in Fig. \ref{fig1}. The corresponding effective charge of
calculations values are $e$=1.832 (W.u.)$^{1/2}$.}

\begin{ruledtabular}
\begin{tabular}{cccc}
\textrm{$L_{i}$}&
\textrm{$L_{f}$}&
\textrm{Expt.$^{a}$}&
\textrm{Calc.$^{b}$}\\
\colrule
$2_{1}^{+}$ & $0_{1}^{+}$ & 217(5) & 217 \\

$2_{2}^{+}$ & $4_{1}^{+}$ & 0.78(4)& 0.79 \\
\quad & $2_{1}^{+}$ & 9.6(6) & 10 \\
\quad & $0_{1}^{+}$ & 5.17(21) & 5.66\\

$2_{3}^{+}$ & $4_{1}^{+}$ & 39(6) & 0.38\\
\quad & $2_{1}^{+}$ & 0.013(13) & 0.18 \\
\quad & $0_{1}^{+}$ & 0.66(8)/0.018(2) & 0.12 \\

$0_{2}^{+}$ & $2_{1}^{+}$ & 2.7(10) & 0.74\\
\quad & $2_{2}^{+}$ & \quad & 1.15\\

$0_{3}^{+}$ & $2_{1}^{+}$ &$<0.83$& 0.84\\

$3_{1}^{+}$ & $4_{1}^{+}$ & 4.8(9) & 6.5\\
\quad & $2_{2}^{+}$ & \quad & 385\\
\quad & $2_{1}^{+}$ & \quad & 10.1\\

$4_{1}^{+}$ & $2_{1}^{+}$ & 312(11) & 306\\

$4_{2}^{+}$ & $6_{1}^{+}$ & 2.01(14) & 2.1\\
\quad & $4_{1}^{+}$ & 11.1(7) & 12.2\\
\quad & $3_{1}^{+}$ & 370(30) & 285\\
\quad & $2_{2}^{+}$ & 138(9) & 126\\
\quad & $2_{1}^{+}$ & 1.98(12) & 2.4\\

$5_{1}^{+}$ & $6_{1}^{+}$ & 12.4(15) & 9.4\\
\quad & $4_{2}^{+}$ & 310(40) & 206\\
\quad & $4_{1}^{+}$ & 8.9(11) & 7.7\\
\quad & $3_{1}^{+}$ & 300(40) & 200\\

$6_{1}^{+}$ & $4_{1}^{+}$ & 0.88(6) & 1.38 \\

$6_{2}^{+}$ & $8_{1}^{+}$ & 1.9(3) & 3.3\\
\quad & $6_{1}^{+}$ & 9.9(7) & 12.6\\
\quad & $5_{1}^{+}$ & \quad & 147\\
\quad & $4_{2}^{+}$ & 225(16) & 240\\
\quad & $4_{1}^{+}$ & 0.88(6) & 1.38\\

$7_{1}^{+}$ & $8_{1}^{+}$ & 8.0(16)& 11.7\\
\quad & $6_{2}^{+}$ & \quad & 117\\
\quad & $6_{1}^{+}$ & 3.4(7) & 6\\
\quad & $5_{1}^{+}$ & 220(40) & 267\\

$8_{1}^{+}$ & $6_{1}^{+}$ & 373(14) & 334 \\

$8_{2}^{+}$ & $10_{1}^{+}$ & 1.5& 4.5\\
\quad & $8_{1}^{+}$ & 8.5(9)& 12.5\\
\quad & $6_{2}^{+}$ & 250(23) & 275\\
\quad & $5_{1}^{+}$ & 0.52(5) & 0.78\\

$9_{1}^{+}$ & $7_{1}^{+}$ & 370(15) & 288 \\

$10_{1}^{+}$ & $8_{1}^{+}$ & 390(17) & 328 \\

$10_{2}^{+}$ & $8_{2}^{+}$ & 290(60)& 278\\
\quad & $8_{1}^{+}$ & 1.5(3)& 0.4\\

$12_{1}^{+}$ & $10_{1}^{+}$ & 372(21) & 318 \\

$14_{1}^{+}$ & $12_{1}^{+}$ & 400(50) & 295 \\

$16_{1}^{+}$ & $14_{1}^{+}$ & 330(18) & 273 \\

\end{tabular}
\end{ruledtabular}
\end{table}

The reduced transition probability $B(E2)$ serves as a critical probe of collective nuclear structure. Table \ref{tab:table3} presents a comparison between experimental and calculation $B(E2)$ values for $^{166}$Er. It can be seen that the SU3-IBM calculations are in good agreement with the majority of experimental results. However, a significant discrepancy is observed for the $B(E2;2_{3}^{+}\rightarrow 4_{1}^{+})$, where the calculated value is much smaller than the experimental result. Interestingly, this anomaly may reflect experimental uncertainty rather than theoretical deficiency. Across the Er isotopic chain, the measured $B(E2;2_{3}^{+}\rightarrow 4_{1}^{+})$ values for  $^{168}$Er and $^{170}$Er are 0.87 W.u. and 1.42 W.u., respectively,  whereas the reported value for $^{166}$Er is anomalously large. This systematic trend suggests that the experimental $B(E2)$ value for $^{166}$Er may be somewhat overestimated. We hope that further experiments can be conducted to validate our assessment. In addition, our model predicts several unmeasured BE(2) values, including $B(E2;0_{2}^{+}\rightarrow2_{2}^{+})$, $B(E2;3_{1}^{+}\rightarrow2_{2}^{+})$ ,$B(E2;3_{1}^{+}\rightarrow2_{1}^{+})$, $B(E2;6_{2}^{+}\rightarrow5_{1}^{+})$, $B(E2;7_{1}^{+}\rightarrow6_{2}^{+})$. These predictions await experimental verification, which can also provide further opportunities to test our model.

\begin{table}
 \renewcommand\arraystretch{1.3}
\caption{\label{tab:table4}%
Quadrupole moments in eb  for some low-lying states in $^{166}$Er and various models.}
\begin{ruledtabular}
\begin{tabular}{cccc}
\textrm{$Q(L_{i})$}&
\textrm{Expt.}&
\textrm{MCSM}&
\textrm{SU3-IBM}\\
\colrule
$Q(2_{1}^{+})$ & -1.9(4) & -2.0 & -2.19 \\

$Q(2_{2}^{+})$ & 2.2(3) & 2.0 & 2.19 \\

$Q(4_{1}^{+})$ & -2.7(9) & -2.5 & -2.79\\

\end{tabular}
\end{ruledtabular}
\end{table}
The spectroscopic quadrupole moment $Q$, serving as a direct measure of nuclear deformation, provides critical insights into shape coexistence and triaxiality. For $^{166}$Er, Table \ref{tab:table4} compare the experimental $Q$ values for the states $2_{1}^{+}$, $2_{2}^{+}$ and $4_{1}^{+}$ with calculations from different theoretical approaches. As shown in Table \ref{tab:table4}, the $Q(2_{1}^{+})$ calculated by the MCSM is more consistent with the experimental results, however, our approach provides superior descriptions of $Q(2_{2}^{+})$ and $Q(4_{1}^{+})$. Overall, our calculations are in excellent agreement with the experimental data, which suggests that the SU3-IBM framework can captures the SU(3) triaxial deformation characteristics of $^{166}$Er comprehensively.

\section{Discussions}
In Ref. \cite{smirnov2000}, the relation between collective rotor parameters $(\beta, \gamma)$ and the SU(3) irrep labels $(\lambda, \mu)$ was given in Fig. 1 for $N=6$. It was emphasized that ``The SU(3) irreps valid for the IBM are only a subset of those which are allowed in the shell model in accordance with the Pauli principle''. Therefore, for a finite boson number $N$, only a limited set of discrete $\gamma$ values can be obtained from Eq. (\ref{eq9}), thereby limiting the ability of the SU(3) symmetry limit to describe the nuclear shape deformations. For instance, the experimental $\gamma$ value of $^{154}$Sm ($N=11$) is $5.0^{\circ}$ and calculated value reported in \cite{otsuka2025} is $3.7^{\circ}$, neither of which corresponds to an allowable SU(3) irrep. However, this discrepancy diminishes as the boson number increases. For $^{166}$Er ($N = 15$), the irrep $(22,4)$ can give a value comparable to that reported in Ref. \cite{otsuka2025}. This improvement arises because larger $N$ value admits more admissible irreps, yielding a denser grid of discrete $\gamma$ values. Nevertheless, the fundamental limitation of the SU(3) symmetry limit persists regardless of boson number $N$.

It is important to emphasize that further adjustment of $\gamma$ value can be realized through the inclusion of the $\hat{n}_{d}$ and $\hat{H}_{D}$ terms. Because these terms break the SU(3) symmetry and lead to the mixing among different irreps, which can make an appropriate correction to the $\gamma$ value via the weighted average in Eq. (\ref{eq11}). This symmetry-breaking mechanism and its quantitative implications for triaxiality will constitute the focus of our subsequent investigations.

\section{Conclusions}
In summary, the triaxial deformation of $^{166}$Er is described within the SU3-IBM framework, where higher-order interactions are included. In this model, all quadrupole deformations are governed by the SU(3) symmetry, which provides a algebraic foundation for describing triaxial nuclear shapes. Under the dominant SU(3) irrep $(\lambda,\mu)=(22,4)$, a triaxiality parameter of $\gamma=9.7^{\circ}$ is obtained. In addition, energy levels, $B(E2)$ values and collective quadrupole moments calculated in our model show excellent agreement with experimental data. This consistent agreement across multiple observables provides strong support for SU(3) triaxial deformation in $^{166}$Er, rather than purely prolate deformation.

\begin{acknowledgments}
This work is supported by the National Natural Science Foundation of China (Grant No. 12104150), Scientific Research Fund of Hunan Provincial Education Department (Grant No. 24B0639).
\end{acknowledgments}


\end{document}